\documentclass[aps,prl,amsmath,amssymb,twocolumn]{revtex4-1}
\usepackage[utf8]{inputenc}
\usepackage{bm}        
\usepackage{verbatim}   
\usepackage{braket}
\usepackage{afterpage}
\usepackage{amsmath}
\usepackage{nth}
\usepackage{physics}
\usepackage{graphicx}

\usepackage{pdfpages}

\newcommand{\des}{\hat{a} }
\newcommand{\cre}{\ensuremath{\hat{a}^{\dagger} } }

\newcommand{\EJ}{\ensuremath{ E_{\text{J}}}}

\newcommand{\im}{\ensuremath{ \text{i} }}

\newcommand{\nlambda}{\ensuremath{\lambda}}

\makeatletter
\AtBeginDocument{\let\LS@rot\@undefined}
\makeatother

\begin{document}

\title{Discrete time translation symmetry breaking in a Josephson junction laser}

\author{Ben Lang, Grace F. Morley and Andrew D. Armour}

\affiliation{School of Physics and Astronomy and Centre for the Mathematics and Theoretical Physics of Quantum Non-Equilibrium Systems, University of Nottingham, Nottingham NG7 2RD, U.K.}

\begin{abstract}
A Josephson junction laser is realised when a microwave cavity is driven by a voltage-biased Josephson junction. Through the ac Josephson effect, a dc voltage generates a periodic drive that acts on the cavity and generates interactions between its modes. A sufficiently strong drive enables processes that down-convert a drive resonant with a high harmonic into photons at the cavity fundamental frequency, breaking the discrete time translation symmetry set by the Josephson frequency. Using a classical model, we determine when and how this transition occurs as a function of the bias voltage and the number of cavity modes. We find that certain combinations of mode number and voltage tend to facilitate the transition which emerges via an instability within a subset of the modes. Despite the complexity of the system, there are cases in which the critical drive strength can be obtained analytically. 
\end{abstract}
\maketitle

 Circuit QED systems are ideally suited to the exploration of nonlinear phenomena, such as frequency conversion, which underlie the breaking of discrete time translational symmetries\,\cite{Gu_2017,Svensson_2017,Sandbo_2020}.  Famously, applying a dc voltage, $V$, to an isolated Josephson junction (JJ) leads to oscillations in the junction phase at the Josephson frequency, $\omega_J=2eV/\hbar$. However, if the junction is embedded in a microwave cavity, interactions between the strongly non-linear junction and cavity modes can trigger oscillations at frequencies below $\omega_J$, breaking the discrete time-translational symmetry (DTTS). The resulting down conversion processes and the properties of the radiation they generate have been widely explored in few mode systems over the last few years, both experimentally\,\cite{Hofheinz_2011,Westig_2017,Peugeot_2021,Menard_2021} and theoretically\,\cite{Leppakangas_2013,Zhang_2017,Gosner_2020,Lang_2021,Arndt_2022}.

 The breaking of a DTTS has been studied extensively elsewhere in the context of many-body systems of coupled oscillators and the symmetry broken time-crystalline phases that emerge\,\cite{Dykman_2018,Yao_2020,Kesler_2020,Kessler_2021, Heugel_2022}. For superconducting circuits, although attention has largely focused on systems where just one or two modes play an important role, the possibilities enabled by utilising multiple modes are attracting increasing attention\,\cite{Chakram_2021}. In particular, recent  experiments revealed strong coherent radiation at the fundamental mode frequency in a JJ-cavity system where $\omega_J$ matched a high overtone of the cavity\,\cite{Cassidy_2017}. In this system, known as a Josephson junction laser (JJL)\,\cite{Cassidy_2017,Steven_2018}, emission at the fundamental mode frequency is believed to result from an interplay between many of the cavity's modes\,\cite{Cassidy_2017}.
 
 \begin{figure}[t]
\includegraphics[scale=0.22]{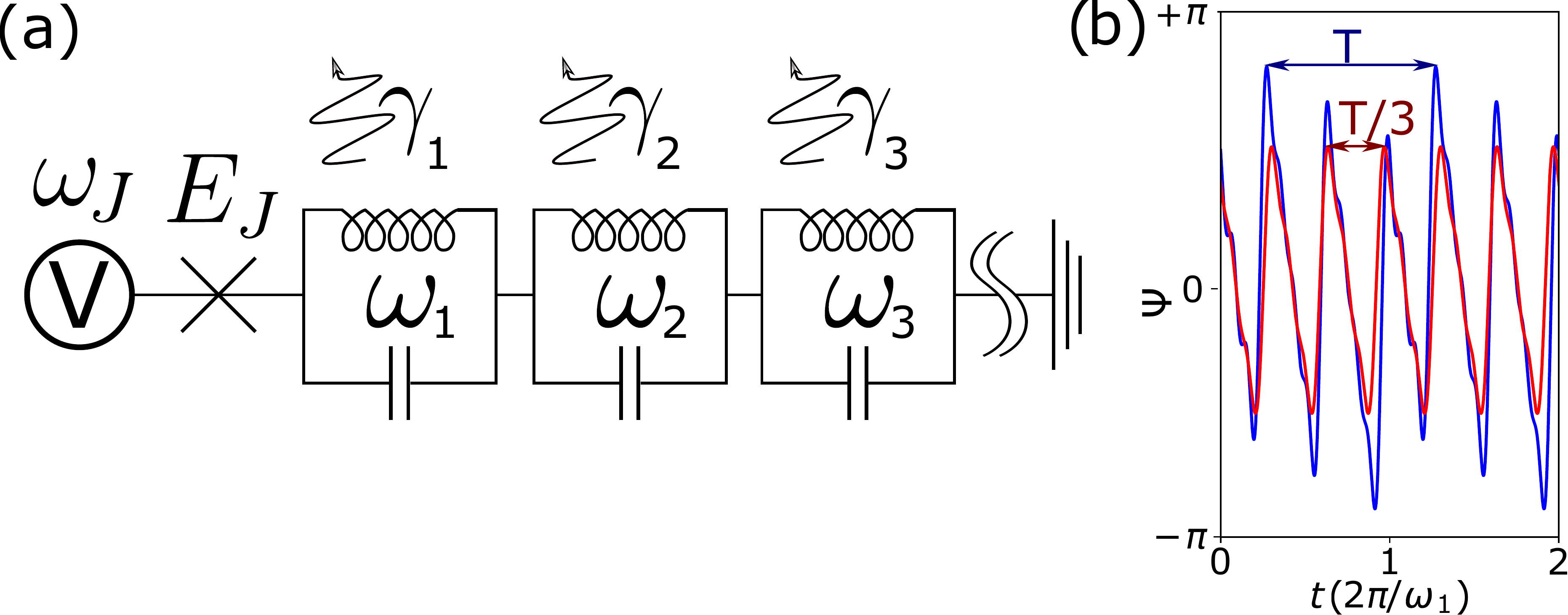}
\caption{(a) Circuit model of the JJL. A JJ with Josephson energy, $E_J$, biased by a voltage, $V$, in series with a microwave cavity modelled as a series of LC oscillators with frequencies $\omega_1,\omega_2,\dots$. Losses from the cavity lead to damping of the modes with rates $\gamma_1, \gamma_2,\dots$.  (b) DTTS breaking transition in the oscillations of the total cavity phase, $\Psi(t)$: above the transition (blue) these have the period of the fundamental mode, $T=2\pi/\omega_1$, but period $2\pi/\omega_J$ below (red). Here, $N=11$, $\omega_J=3\omega_1$ and the origin of the time axis has been displaced to display the long-time behavior clearly.}
\label{circuit}
\end{figure}
Inspired by the JJL experiments, in this Letter we analyse a simplified theoretical model of a voltage-biased JJ in series with a cavity, described as a set of $N$ harmonic modes (see Fig \ref{circuit}a). 
We explore the many-body discrete time-translational symmetry breaking (DTTSB) transition which manifests as a change in the periodicity of the cavity oscillations (see Fig \ref{circuit}b): from $2\pi/\omega_J$ to a larger value set by the period of the fundamental mode. Numerical modelling has highlighted the crucial role played by the multiple modes supported by the cavity\,\cite{Cassidy_2017} and a solution obtained analytically for the time-crystal like symmetry broken regime\,\cite{Steven_2018}. However, the fundamental questions we seek to answer of when and how the transition occurs have not been addressed. 

The JJ leads to a Hamiltonian which ostensibly generates all-to-all couplings between cavity modes, but most of these couplings are irrelevant as they are far off-resonant. Adopting a rotating wave approximation and a coherent state ansatz, we find that prior to the transition the modes fall into noninteracting subspaces with different symmetry properties. This division into subspaces means that the location of the DTTSB transition proves surprisingly dependent on the number of cavity modes. Relatively low transition thresholds are associated with cases where the transition is continuous and we demonstrate that in such cases it initially arises from an instability affecting modes in a single subspace. Remarkably, closed form expressions for the critical coupling can be obtained in some cases.

\emph{Model System}.---The JJL can be modelled as a set of harmonic cavity modes\,\footnote{Note that the description in terms of cavity modes can be reformulated in terms of the continuum limit of a chain of locally coupled lumped element $LC$ oscillators.} in series with a JJ to which a dc voltage bias $V$ is applied\,\cite{Armour_2013,Armour_2015,Chen_2014,Cassidy_2017,Steven_2018}. The phase across the JJ is controlled by the voltage and contributions from each of the modes leading to a  Hamiltonian\,\cite{Armour_2013,Armour_2015,Trif_2015,Hofer_2016,Dambach_2017}
\begin{equation}
\hat{H}(t)= \sum_{n}  \hbar \omega_{n} \cre_{n} \des_{n} - \EJ \text{cos} \left[ \omega_{\text{J}} t + \sum_{n} \Delta_{n} ( \cre_{n} + \des_{n} ) \right],
\label{time_dependent_hamiltonian}
\end{equation}
where $E_J$ is the Josephson energy of the junction, $\des_n$ is the raising operator for the $n$th cavity mode with frequency $\omega_n$,  and $\Delta_{n}= \sqrt{2e^2/(\hbar C\omega_n)}$ is the corresponding strength of the zero point flux fluctuations (in units of the flux quantum) with $C$ the cavity capacitance\,\cite{Armour_2013}. The Hamiltonian possesses the DTTS, $\hat{H}(t)=\hat{H}(t+T_J)$ with period $T_J=2\pi/\omega_J$.

The explicit time dependence of the cosine term in the Hamiltonian \eqref{time_dependent_hamiltonian} acts as a nonlinear drive whose strength can be tuned by varying $E_J$\,\cite{Armour_2013,Cassidy_2017}. This term is balanced by dissipation, since photons can leak out of the cavity into its surroundings. Assuming zero temperature for simplicity, the dissipation can be described via a standard Lindblad master equation\,\cite{Carmichael_book,Kubala_2015,Armour_2015,Trif_2015,Hofer_2016} 
\begin{equation} 
\dot{\rho} = -\frac{\im}{\hbar}[\hat{H}(t), \rho]
+ \sum_{n} \frac{\gamma_{n}}{2} ( 2 \des_{n} \rho \cre_{n}  - \cre_{n} \des_{n} \rho - \rho \cre_{n} \des_{n} ),
\label{Master_equation}
\end{equation}
with $\gamma_n$ the loss rate for mode $n$.

In the following, we assume a hard cut-off in the number of modes, $N$, together with an idealised cavity spectrum,  $\omega_n=n\omega_1$, where $\omega_1$ is the fundamental mode frequency, and a constant loss rate $\gamma_n=\gamma$\,\cite{Cassidy_2017,Steven_2018}.  In practice, deviations in the mode frequencies and changes to the damping rate will eventually become important as the mode index is increased\,\cite{Armour_2013,Goppl_2008}, leading to an effective decoupling of high frequency modes, but how this occurs will depend on precisely how the JJL is engineered. Rather than concentrate on a single specific realisation, our use of a hard cut-off  instead treats $N$ as a parameter, allowing us to focus on exploring how it affects the location and character of the DTTSB transition. 


\emph{Classical Dynamics}.---We analyse the classical dynamics of the system using a coherent state ansatz to derive an approximate set of coupled classical equations for the mode amplitudes\,\cite{Wood_2021}. We assume that each mode is in a coherent state, $\rho_{\alpha}=\bigotimes_{n=1}^N|\alpha_n\rangle\langle\alpha_n|$, with $\alpha_n$ a complex time-dependent amplitude\,\cite{Armour_2013,Wood_2021}. This approach is expected to become increasingly accurate as the value of $\Delta_1$ is reduced, since $\Delta_1\rightarrow 0$ constitutes the classical limit for the system\,\cite{Armour_2017}, but it also provides a framework for developing a quantum analysis\,\cite{Armour_2013}.

The corresponding classical Hamiltonian $\mathcal{H}(t, \vec{\alpha})$ is obtained from its quantum counterpart [Eq.\ \ref{time_dependent_hamiltonian}] by making the replacements\,\cite{supplement}  $\hat{a}_n^{(\dagger)}\rightarrow \alpha_n^{(*)}$ and $E_J\rightarrow\tilde{E}_J={E}_J{\rm{exp}}[{-\sum_{n=1}^N}\Delta_n^2]$ (this arises when normal-ordering is performed)\,\cite{Grabert_1998,Armour_2013,Gramich_2013,Leger_2019,Peugeot_2021}. Combined with Eq.\ \eqref{Master_equation}, this leads to dynamical equations for the amplitudes of the form
\begin{equation}
\dot{\alpha}_n = -\left(\im\omega_n+\frac{\gamma}{2}\right) \alpha_n- \frac{\im\tilde{E}_J\Delta_n}{\hbar} \sin\left(\omega_Jt+\Psi\right),
\label{RWA_equation_of_motion}
\end{equation}    
where $\Psi=2\sum_n\Delta_n{\rm{Re}}[\alpha_n]$ is the total phase across all of the modes\,\cite{Steven_2018}.

\begin{figure}[t].
\includegraphics[scale=0.44]{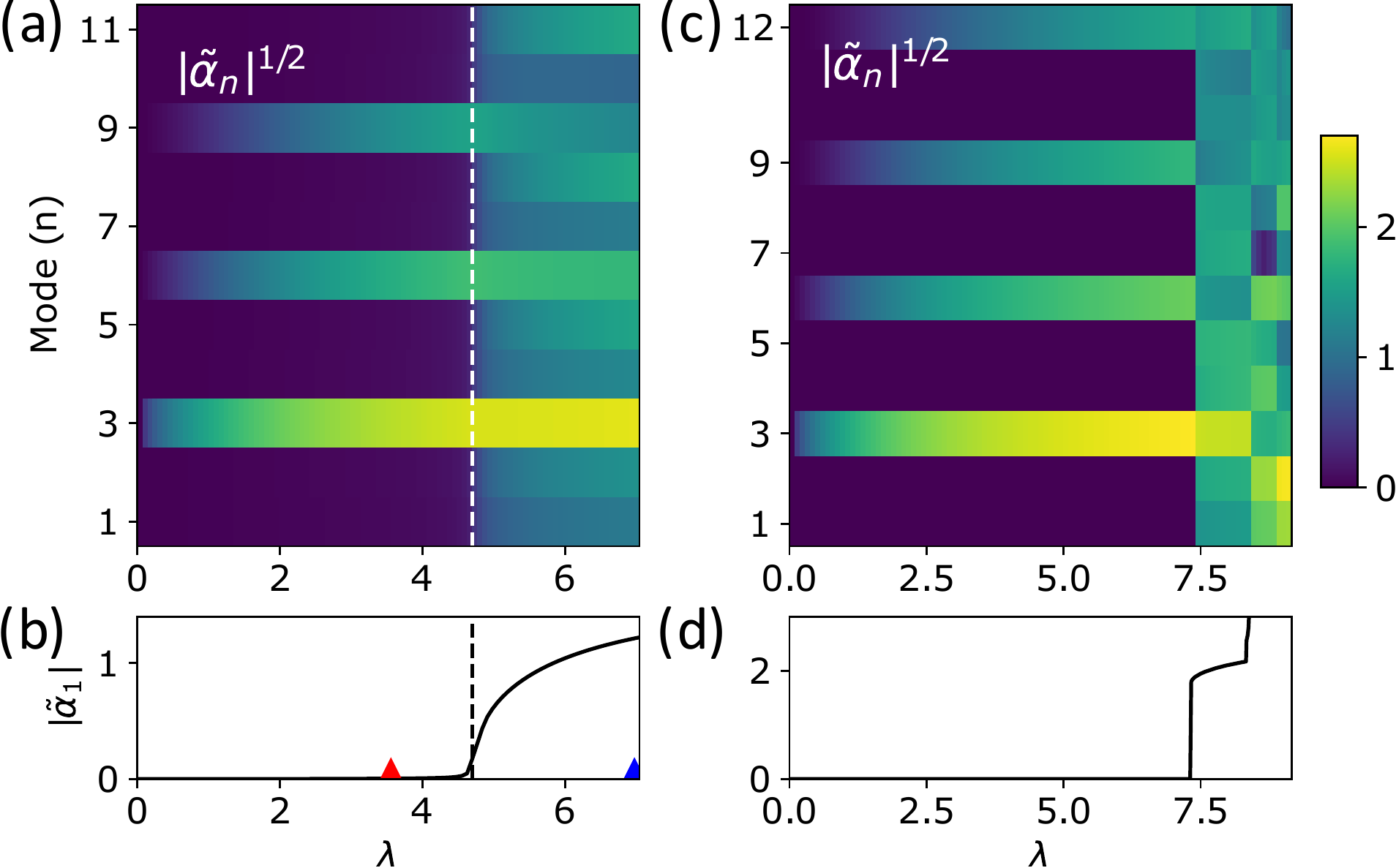}
\caption{Long-time behavior of $\tilde{\alpha}_n$, the component of $\alpha_n$ oscillating at $\omega_n$, as a function of $\lambda$ with $p=3$ for $N=11$ (a,b), a continuous transition, and  $N=12$ (c,d), a discontinuous transition. In both cases $\Delta_1 = 0.2$ and $\gamma = \omega_1 / 100$. (a, c) Show $|\tilde{\alpha}_n|$ for all modes, (b,d) show just the fundamental, $|\tilde{\alpha}_1|$ which serves as an order parameter for the transition. See \cite{supplement} for details of the numerical method. Red and blue symbols in (b) indicate the $\lambda$ values illustrated in Fig.\ \ref{circuit}(b). The vertical dashed lines show the threshold predicted using linear stability analysis (discussed in the text below).
}
\label{class_sims}
\end{figure}
%

To identify the transition, the equations for the amplitudes are integrated numerically. We focus on the case where the drive is resonant with a higher harmonic $\omega_J=p\omega_1$ ($p=2,3,\dots$) and vary the dimensionless drive strength\,\cite{Steven_2018} $\nlambda=\tilde{E}_J\Delta_1^2/\hbar\gamma$. At very weak drives, excitation starts in the resonant mode ($\omega_p=p\omega_1=\omega_J$), but the cosine term in the Hamiltonian (\ref{time_dependent_hamiltonian}) upconverts these oscillations into effective drives at $m\omega_J$ (with $m$ a positive integer), progressively exciting the resonant harmonics---all those modes with frequencies matching $m\omega_J$. The transition occurs when the excitation spreads beyond the resonant harmonics, it is seen clearly in the behavior of $\Psi$ which changes from a sawtooth oscillation of period $T_J=2\pi/\omega_J$ in the symmetry preserving phase to one with period $T_1=2\pi/\omega_1$ when the symmetry is broken (Fig. \ref{circuit}b). 

Tracing the response of the individual mode amplitudes reveals that the transition can be either continuous or discontinuous (see Fig. \ref{class_sims}), depending on the precise relationship between $N$ and $p$. Surprisingly, Fig. \ref{class_sims} shows that changing $N$ by just one can trigger a change from continuous to discontinuous transition accompanied by a large increase in the critical coupling.   

\emph{Continuous Transitions}.---To understand the connections between the characteristics of the transition and the properties of the system, we now focus on analysing cases where a continuous transition occurs.  In such cases important simplifications can be made which allow approximate analytic methods to be employed so that the transition can be located without the need for numerical integration.

The first simplification is obtained by making a rotating wave approximation (RWA)\,\cite{Armour_2013,Gramich_2013,Wood_2021,Peugeot_2021}. Assuming $\omega_J=p\omega_1+\delta$, with $p$ an integer greater than unity and $\delta$ a small detuning, we move to a rotating frame by applying the transformation
\begin{equation}
\hat{U}(t) = \exp(\im \sum_{n=1}^{N} n (\omega_J / p)  \hspace{0.2pc} \cre_{n} \des_{n} \hspace{0.2pc} t).
\label{unitary}
\end{equation}
The RWA is made by discarding non-resonant processes so that each mode is assumed to oscillate only at is own frequency\footnote{The RWA is expected to be a good approximation on-resonance provided $\gamma \ll \omega_n$ and $\EJ \Delta_n \ll \omega_n$ for all $n$\,\cite{Wood_2021}.}. The resulting Hamiltonian, can be written as
\begin{eqnarray}
\hat{H}_{\text{RWA}} &=& \sum_{n=1}^N \hbar \delta_{n} \cre_{n} \des_{n}-\frac{\tilde{E}_J}{2} \left[  Z_{p}^{\{N\}}(  \vec{\hat{x}} ) + {\rm{h.c}}\right]
\label{HRWA}
\end{eqnarray}
where where $\delta_n=-(n/p)\delta$,
 $\vec{\hat{x}}=(\hat{x}_1,\dots,\hat{x}_N)$ with $\hat{x}_n=2\im \Delta_n \des_n$, and we have defined
\begin{equation}
Z_{p}^{\{N\}} (\vec{\hat{x}}) =  : \int_{-\pi}^\pi \frac{dt}{2\pi} \exp[ \sum_{n=1}^N \frac12 \left(\hat{x}_n {\rm{e}}^{\im n t} - {\rm{h.c.}} \right) -\im p t ]:.
\label{NDZint}
\end{equation}
Here colons imply normal ordering and $\left\{N\right\}=1,2,\dots,N$.
Note that the $Z$-functions\,\cite{supplement,Wood_2021} [defined via Eq.\ (\ref{NDZint})] are multi-dimensional generalizations of Bessel functions, analytically continued for complex arguments\,\cite{Dattoli1998, Korsch2008}.

Setting the detuning to zero for simplicity, the DTTS with period $2\pi/\omega_J=2\pi/(p\omega_1)$  manifests as a discrete rotational symmetry in the rotating frame\,\cite{Guo_2013,Guo_2020,Lang_2021}
\begin{equation}
\mathcal{R}( 2\pi / p) \hat{H}_{\text{RWA}} = \hat{H}_{\text{RWA}},
\end{equation}
where  $\mathcal{R}(\theta) \bullet \, = \hat{r}(\theta) \bullet \hat{r}^{\dagger}(\theta)$ with the operator $\hat{r}(\theta) = \exp( \im \theta \sum n \cre_n \des_n )$ rotating a state by an angle $\theta$. The modes can be grouped according to their eigenvalue $\exp( k \im 2\pi/p )$ when acted on by $\mathcal{R}(2\pi/p)$. The resonant harmonics all have $k=0$  the other modes have $k\neq 0$ and fall into sets defined by
\begin{equation}
\vec{s}_k= ( \des_k, \cre_{p-k}, \des_{p+k}, \cre_{2p-k}, \des_{2p+k}, \cre_{3p-k},... ).
\label{sk_vec}
\end{equation}
For odd (even) $p$ there are $(p-1)/2$, ($p/2$) distinct sets\,\cite{supplement}.

 To obtain critical couplings for continuous transitions in the classical regime, we employ the corresponding coherent state ansatz Hamiltonian obtained from Eq.\ \ref{HRWA}, $\mathcal{H}_{\text{RWA}}$\,\cite{supplement}. The fixed point corresponding to the symmetry conserving solution is found, and the critical drive strength, $\lambda_{\text{c}}$, at which this point becomes unstable is identified using linear stability analysis, a detailed description of the method is given in \cite{supplement}. Comparison with numerical integration (see Fig.\ \ref{class_sims}b) demonstrates that this approach can identify the critical coupling  accurately.
 
 Despite the potential complexity of the problem, the linear stability analysis  proves tractable for a wide choice of $N$ and $p$. The fact that we only need to find the fixed points below the transition represents a significant simplification since in this regime only a sub-set of the modes (the resonant harmonics) have non-zero amplitudes, reducing the effective dimensionality of the problem. Furthermore, once obtained,  the fixed point amplitudes for a given set of resonant harmonics can be applied to any combination of ($p$, $N$) with the same number of resonant harmonics via a simple scaling\,\cite{supplement}. We located fixed points for sets of up to nine resonant harmonics which is sufficient for all combinations of $N$ and $p$ such that $N < 10p$. Finally, the special function form of the Hamiltonian [see Eq.\ (\ref{HRWA})] is readily differentiated, facilitating the calculation of the Jacobian matrix used for the stability analysis\,\cite{Wood_2021,supplement}. 

 As well as yielding values of the critical couplings for continuous transitions, our approach also reveals {\emph{how}} the transition occurs. Below the transition, the Jacobian is block diagonal\,\cite{supplement}: couplings only occur between modes with the same eigenvalue  under discrete rotation, $k$. The continuous transition emerges as an instability of the zero amplitude fixed point within just one of the $k\neq 0$ blocks. Numerical integration shows that this instability then spreads progressively to modes with different $k$-values. One can think of the blocks like adjacent dominoes: the stability of each domino is independent of the others, but an instability and subsequent symmetry breaking in one spreads to the others.

\begin{figure}
\includegraphics[scale=0.54]{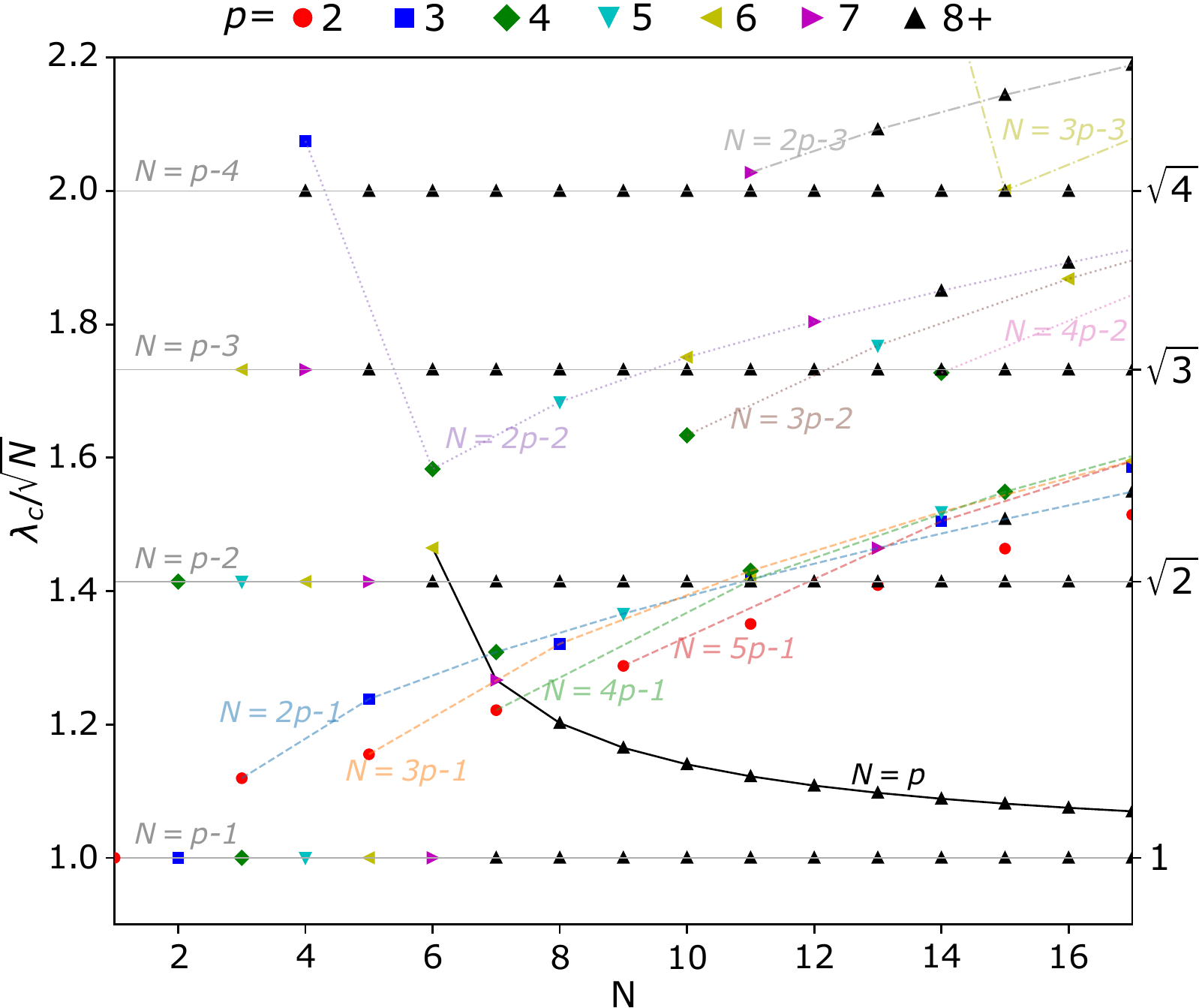}
\caption{Critical drive strength, $\lambda_{\text{c}}$ at which a continuous transition takes place for a range of $N$ and $p$ values, calculated using linear stability analysis\,\cite{supplement}. Lines connect points with a specific relationship between $N$ and $p$ (labelled in each case).}
\label{Cont_Thresholds}
\end{figure}

Figure \ref{Cont_Thresholds} shows how the threshold for a continuous transition depends on both $N$ and $p$. The basic trend is of a critical drive strength that increases with mode number $N$ (note that $\lambda_c/\sqrt{N}$ is plotted).  Even though all of the modes are coupled to the Josephson drive, only one is resonant, hence it is not surprising that the transition typically becomes harder to reach as more modes are added. However, the nonlinear couplings between the modes enable a complex range of frequency conversion processes mediated by the drive, with new ones enabled with each mode added. Understanding these processes is key to understanding the complex interplay between the behavior of $\lambda_c$ and the precise relationship between $N$ and $p$ revealed in Fig.\ \ref{Cont_Thresholds}. 

Qualitatively different trends emerge for the sets $p/2\le N<p$, $N=p$ and $N>p$,  which we consider in turn. Note that no continuous transition is possible for $N<p/2$. In this case, the lowest order processes are cubic or higher (creating 3 or more photons), and do not lead to continuous transitions \cite{Zhang_2017, Lang_2021}.

For $p/2<N<p$ an instability occurs via a parametric down conversion process in which the Josephson drive excites two modes whose frequencies sum to $\omega_J=p\omega_1$\footnote
{When $p=2N$ the $N$th mode acts as a degenerate parametric amplifier whose instability drives a continuous transition}. The different blocks in the Jacobian, corresponding to different values of $k$, couple the pairs of modes $\omega_k$ and $\omega_{p-k}$, in each case the interaction is equivalent to that in a non-degenerate parametric amplifier. The blocks differ only by factors of $\Delta_n=\Delta_1/\sqrt{n}$ (this scaling follows directly from the fact that the modes are harmonics of a microwave cavity with $\omega_n=n\omega_1$\,\cite{Goppl_2008}), with the parametric terms in block $k$ proportional to $\Delta_k \Delta_{p-k}$. Since $\Delta_1 \Delta_{p-1} > \Delta_2 \Delta_{p-2} > \Delta_3 \Delta_{p-3}$... the most strongly excited parametric process involves the last mode added, $\omega_N$, and one finds $\nlambda_{\text{c}} = \sqrt{N(p-N)}$ \,\cite{supplement}, leading to  the horizontal lines in Fig.\ \ref{Cont_Thresholds}. They form a ladder: keeping $p$ fixed and increasing N moves us down the rungs.

The second regime is $N = p$, where the system now possesses a resonant mode and the critical coupling follows the solid black curve sloping downward in Fig. \ref{Cont_Thresholds}. The presence of a resonant mode impedes the transition by diverting energy away from the dominant parametric terms that now have a reduced coupling strength in the Jacobian. The critical drive is found to be\,\cite{supplement} $\nlambda_{\text{c}} = \sqrt{p-1}/[2 J_1'(z_p)]$, with $J_1(x)$ a Bessel function and  $z_p$ the fixed point amplitude of the resonant mode multiplied by $2\Delta_p$, which satisfies\,\cite{Armour_2013} $z_p^2 = (4\lambda/p) J_1(z_p)$.

The impeding effect of the resonant mode is strongest at low $p$ values. For $p=N < 6$ no continuous transition occurs at all, but as $p$ is increased the impact of the resonant mode gets weaker (since for a given $\lambda$, $z_p$ gets smaller for larger $p$,  $J_1'(z_p)$ increases with $p$) and a simpler approximate relation can be derived\,\cite{supplement}, $\nlambda_{\text{c}} \approx p \sqrt{p-1}/ (p-3/2)$, for large $p$. Eventually $\lambda_c$ tends to $\sqrt{N}$, matching the critical drive for $N=p-1$, in the $p\rightarrow \infty$ limit.

Finally we consider cases where $N>p$. The number of resonant harmonics grows with $N$, making it progressively more difficult to obtain the below-transition fixed point. Further complexity is introduced by an increase in the number of modes within each block beyond two, enabling beam-splitter type interactions in which modes with frequency {\emph{difference}} a multiple of $\omega_J=p\omega_1$ exchange energy with each other mediated by the drive and the resonant harmonics. Nevertheless, the blocks with different $k$ remain fundamentally the same, differing only by factors of $\Delta_n$ and by how they are truncated by the finite mode number.

In this regime it is helpful to think about the processes enabled by the last mode added: relatively low values of $\lambda_c$ are found where this mode enables additional parametric processes in which two modes can jointly be excited. The best example occurs when $N=qp - 1$ for $q=2,3$..., in each such case the $N$th mode is added to the $k=1$ space, and is parametrically coupled to the fundamental: these modes can be excited together by combinations of the drive and the resonant harmonics ($\omega_1+\omega_N=qp\omega_1$). Similar effects arise when $N=qp - 2$,  $N=qp-3$, etc., with the $N$th mode adding a new parametric process to the $k=2,3,\dots$ space. But as  $\Delta_1 \Delta_{qp-1} >\Delta_2 \Delta_{qp-2} > \Delta_3 \Delta_{qp-3}$ the new process added gets weaker with increasing $k$ and the corresponding value of $\lambda_c$ increases (similar to the ladder seen for $N<p$, see Fig.\ \ref{Cont_Thresholds}). From this it follows that for $N=qp-k$ and $k\leq(p/2)$ any DTTSB is in subspace $k$\,\cite{supplement}.



\emph{Discontinuous Transitions}.--- 
We now examine the drive at which the DTTS breaking transition actually occurs in a given system, $\lambda_{\text{sb}}$, which need not necessarily correspond to $\lambda_c$ since a {\emph{discontinuous}} transition is also possible (see Fig. \ref{class_sims}). 
Discontinuous transitions have to be identified via numerical integration by tracing how the symmetry broken state evolves as the drive strength is lowered progressively until it becomes unstable\,\cite{supplement}.

\begin{figure}
\includegraphics[scale=0.65]{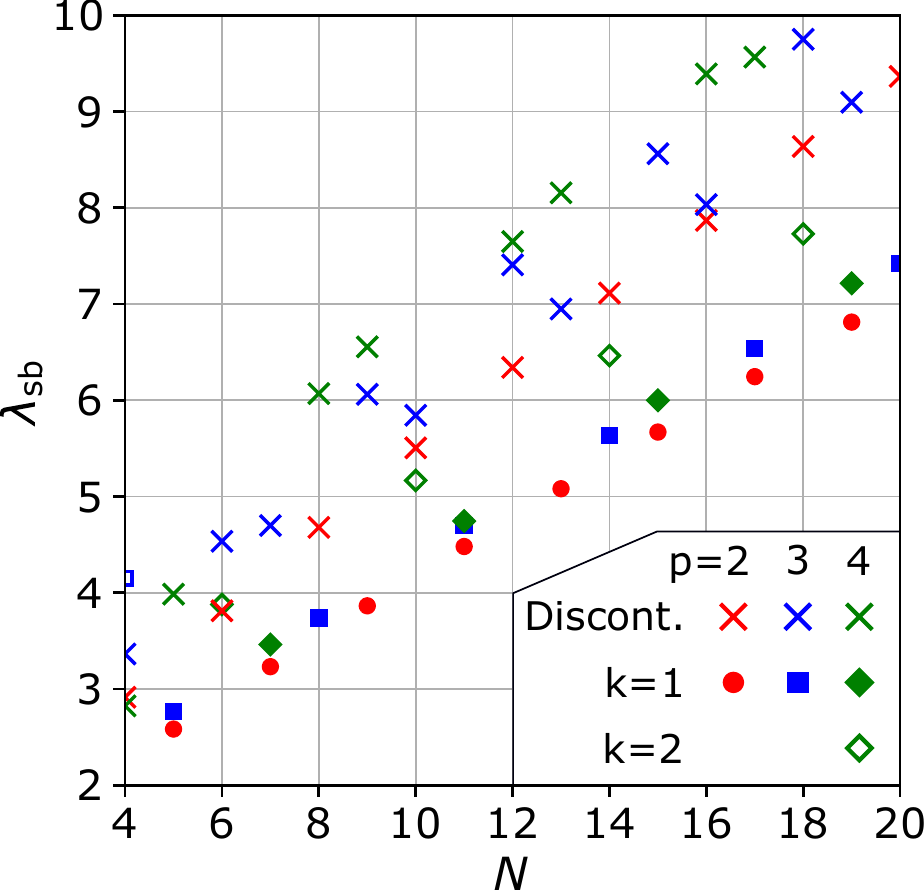}
\caption{Drive at which a DTTS breaking transition occurs in a given system, $\lambda_{\text{sb}}$ ,as a function of $N$ for $p=2,3,4$. Crosses: discontinuous transitions. Solid (hollow) points: continuous transitions originating within  the $k=1$ ($k=2$) block.}
\label{Discont_Thresholds}
\end{figure}

Combining numerical integration data with the critical drives we previously obtained for continuous transitions, Fig.\ \ref{Discont_Thresholds}
shows the value of $\lambda_{\text{sb}}$ as a function of $N$ for $p=2$, $3$ and $4$. The basic message is that the $N$ and $p$ values that we identified as leading to a relatively small  $\lambda_{\text{c}}$ (i.e. within the range shown in Fig.\ \ref{Cont_Thresholds}), a continuous transition does indeed occur before any discontinuous one is reached, but where a continuous transition requires a relatively large drive strength it will generally  be forestalled by a discontinuous one. Specifically, the relatively low values of $\lambda_c$ that arise for $N=qp-1$ (with $q=2,3,\dots$) facilitate continuous transitions at a value of $\lambda_{\rm{sb}}$ significantly lower than neighboring discontinuous transitions in each case. These are in fact the only continuous transitions that occur for $p=2$ and $p=3$, which only possess blocks with $k=0$ and $1$. For $p=4$, which also has a $k=2$ block, continuous transitions occur when $N=4q-2$ (triggered within the $k=2$ block) as well as when $N=4q-1$ (triggered in the $k=1$ block).   


\emph{Conclusion}.---The transition in a many-mode JJ-cavity system that breaks the DTTS of the Hamiltonian occurs either continuously or discontinuously as a function of the system parameters. We developed an approximate classical description of the system dynamics that enabled us to determine when a continuous transition occurs efficiently, leading to simple analytic expressions in some cases. We also examined exactly how these transitions occur: they involve an instability within a particular subset of modes linked by a symmetry property. This uncovered the existence of `magic' combinations of Josephson frequency and mode number which facilitate continuous transitions at relatively low drive strengths. In systems that do not match one of these combinations, the DTTS breaking transition is instead found to occur discontinuously at a relatively high drive strength. 

Beyond the insights that our analysis gives into the complex and unusual classical dynamics of the JJL, it opens the way for a quantum analysis of the system which would reveal the extent to which couplings within each block might give rise to patterns of many-body entanglement and the extent to which quantum effects influence the location and nature of the DTTS breaking transition.  Finally, we note that our analysis could be adapted to describe a range of other nonlinear multi-mode circuit-QED systems driven in different ways, including e.g.\ by an ac flux bias\,\cite{Svensson_2017}.


{\it{Acknowledgement}}: The work was supported by a Leverhulme Trust Research Project Grant (RBG-2018-213). 

\bibliography{references}

\newpage
\clearpage


\section*{Supplementary material}

# Discrete time translation symmetry breaking in a Josephson junction laser: Supplemental Material

Ben Lang, Grace F. Morley and Andrew D. Armour
School of Physics and Astronomy and Centre for the Mathematics
and Theoretical Physics of Quantum Non-Equilibrium Systems,
University of Nottingham, Nottingham NG7 2RD, U.K.

## I. NUMERICAL INTEGRATION OF CLASSICAL EQUATIONS

Classical equations of motion are obtained using the coherent state ansatz. This involves the the assumption that all of the modes are always in coherent states with corresponding time-dependent eigenvalues, $\{\alpha_l\}$. Normally-ordering the Hamiltonian, one finds that these obey the following set of equations of motion [1]

$$\dot{\alpha}_l(t) = -\frac{\mathrm{i}}{\hbar}\frac{\partial \mathcal{H}(t,\vec{\alpha})}{\partial \alpha_l^*} - \frac{\gamma}{2}\alpha_l(t), \tag{S1}$$

with the classical Hamiltonian

$$\mathcal{H}(t,\vec{\alpha}) = \sum_{n=1}^{N}\hbar\omega_n|\alpha_n|^2 - \tilde{E}_J\cos\left[\omega_\mathrm{J} t + \sum_{n=1}^{N}2\Delta_n\mathrm{Re}[\alpha_n]\right], \tag{S2}$$

where $\tilde{E}_J = E_\mathrm{J}\exp\left[-\sum_{n=1}^{N}\Delta_n^2/2\right]$.

The coupled system of equations (S1) is solved numerically, using standard routines, to study the transition to the symmetry broken state. The principal difficulty lies in identifying the location of discontinuous transitions as in such cases the emergence of a (stable) symmetry broken fixed point is accompanied by the coexistence of a stable symmetry preserving one. Our approach is to start by identifying the symmetry broken fixed point well above the transition and seeking to follow it until it eventually disappears as the drive strength is reduced. Specifically, we begin with a random initial condition and a large $\lambda$ value and evolve in time until the system has has relaxed to a steady state with period $T = 2\pi/\omega_1$. This final state is then used as the initial condition for the next calculation at a slightly lower $\lambda$, and so on progressively until the period $T$ disappears. Note that we extract the component of $\alpha_n$ oscillating at $\omega_n$ by evaluating $\tilde{\alpha}_n = \int_0^{10T}(dt/10T)\alpha_n(t)\exp(\mathrm{i}\omega_n t)$.

The discontinuous transition thresholds shown in Fig. 3 of the main text are found using this approach. As there were many such calculations we used additional techniques to save computer resources. This consisted of a first-sweep calculation with large $\lambda$ increments and a relatively large $\gamma/\omega_1 = 3\times 10^{-2}$ to accelerate the convergence to a steady state. The point where the non-resonant modes become inactive with lowering $\lambda$ is identified in this first sweep and a second sweep is carried out in this region. This second sweep used $\gamma/\omega_1 = 1\times 10^{-2}$ and made use of a steady state from the previous sweep as the initial condition for its highest $\lambda$, then sweeping down in small steps.

## II. ANALYSIS OF CONTINUOUS TRANSITIONS

In this section we detail the methods used to identify the locations of continuous transitions. The first steps (described in the main text) are a unitary transformation of the Hamiltonian (moving to a rotating frame) and a rotating wave approximation (RWA). We then obtain the classical equations of motion by applying the coherent state ansatz to the RWA Hamiltonian [Eq. 5]. Assuming that the detunings are all set to zero, this leads to the following equations of motion [1]:

$$\dot{\alpha}_l = -\frac{\mathrm{i}}{\hbar}\frac{\partial \mathcal{H}_\mathrm{RWA}(\vec{\alpha})}{\partial \alpha_l^*} - \frac{\gamma}{2}\alpha_l(t) \tag{S3}$$

$$= -\frac{\Delta_l\lambda\gamma}{2\Delta_1^2}\left[Z_{p+l}^{\{N\}}(2\mathrm{i}\vec{\Delta}\circ\vec{\alpha}) - Z_{p-l}^{\{N\}}(-2\mathrm{i}\vec{\Delta}\circ\vec{\alpha}^*)\right] - \frac{\gamma}{2}\alpha_l. \tag{S4}$$

where the classical RWA Hamiltonian takes the form

$$\mathcal{H}_\mathrm{RWA}(\vec{\alpha}) = -\frac{\tilde{E}_J}{2}\left[Z_p^{\{N\}}(2\mathrm{i}\vec{\Delta}\circ\vec{\alpha}) + \mathrm{h.c.}\right]. \tag{S5}$$

Here we have adopted the vector notation so that $\vec{\Delta} = (\Delta_1, \Delta_2, \ldots, \Delta_N)$ and $\vec{\alpha} = (\alpha_1, \alpha_2, \ldots, \alpha_N)$; the symbol $\circ$ indicates an element-wise product, i.e. $\vec{c} = \vec{a} \circ \vec{b}$ implies that $c_n = a_n b_n$. Note that the Z-functions are defined in the main text [Eq. (6)] and their properties (including differentiation) are discussed in detail in Ref. 1.

To obtain the critical drive strength for a continuous transition we need to identify the non-zero amplitudes at the symmetry preserving fixed point (i.e. those of the resonant harmonics) and then determine the eigenvalues of the associated Jacobian. We address these in turn below (Secs. II A and II B), after which we turn to examine the different uncoupled symmetry blocks that describe the system's dynamics in this regime (Sec. II C).

## A. Symmetry Preserving Fixed Point

Prior to a continuous transition, only resonant harmonics (i.e. modes at integer multiples of the drive frequency) are excited, the others have zero amplitude. This makes finding a fixed point much easier as we can neglect all of the inactive modes, working instead within a reduced mode space that includes just the resonant harmonics. When $pM \leq N < p(M+1)$ the $N$ modes contain $M$ resonant harmonics, this means we can simply replace $Z_p^{1,2,\ldots,N}$ by $Z_p^{p,2p,\ldots,Mp}$ in Eq. S4.

Furthermore, a simple re-scaling allows us to map the problem onto one which only depends on the number of resonant harmonics, $M$, but is independent of the choice of $p$. The mapping relabels the mode index for the resonant harmonics $n \to n'$, where $n'$ runs from 1 to $M$, and $p \to p' = 1$, using the identity [1] $Z_p^{p,2p,\ldots,Mp} = Z_1^{1,2,\ldots,M}$. The values of the zero-point uncertainties and mode amplitudes also need to be rescaled, since mode $n'$ in the reduced space refers to mode $n = pn'$ in the full space $\Delta'_{n'} = \Delta_{pn'} = \Delta_n/\sqrt{p}$. The mode amplitudes undergo the opposite scaling so that $\Delta'_{n'}\alpha'_{n'} = \Delta_{pn'}\alpha_{pn'}$. The dimensionless drive, $\lambda$, remains unchanged.

The point of this re-scaling is that once a solution is found for a given value of $M$, it can be used to describe the below-transition mode amplitudes of any situation with the same number of resonant harmonics (i.e. where $pM \leq N < p(M+1)$). We carried out 9 fixed point calculations [1], with $M$ from 1 to 9. These provide the fixed points needed to describe the below-transition dynamics for any $N$, $p$ such that $N < 10p$. In finding these fixed points we are helped by the fact that for $\lambda = 0$ all modes have the zero amplitude solution and that at each step, as $\lambda$ is incremented higher, the new fixed point is located near the previous one, which was used as the solver's initial guess. At each stage we located the fixed points by evaluating the Z-functions using the integral definitions [Eq. (5)] and varying the values of the amplitudes in small increments away from the previous fixed point.

The behavior of the resonant harmonic fixed points has two stages: First, a phase-locked regime in which the complex mode amplitudes $\alpha'_n = A'_n \exp(i\theta'_n)$ have phases that remain fixed at $\theta_m = (1-m)\pi/2$. Here only the moduli $A'_n = |\alpha'_n|$ change with $\lambda$. At high $\lambda$ this regime ends in a bifurcation, after which the mode phases change with $\lambda$ and the problem becomes more difficult to solve. The $M=1$ case of this bifurcation is described in Ref.[2]. We tracked the fixed point amplitudes as a function of $\lambda$ for each of the resonant harmonics up to this bifurcation.

## B. Jacobian

The stability of the symmetry preserving fixed point is determined by the corresponding Jacobian matrix, $\overline{J}$. Casting the equations of motion (S4) in the compact form $\dot{\vec{\alpha}} = \vec{f}(\vec{\alpha}, \vec{\Delta}, \lambda, \gamma)$, the Jacobian is defined as

$$\overline{J}^T = \left( \frac{\partial}{\partial \alpha_1} \ \frac{\partial}{\partial \alpha_1^*} \ \frac{\partial}{\partial \alpha_2} \ \frac{\partial}{\partial \alpha_2^*} \ldots \right)^T \left( f_1 \ f_1^* \ f_2 \ f_2^* \ldots \right). \tag{S6}$$

The necessary derivatives can be evaluated straightforwardly [1], and we find

$$\frac{\partial f_j}{\partial \alpha_l} = -i\Delta_j \Delta_l G_{j-l} - \frac{\gamma}{2}\delta_{l,j}, \tag{S7}$$

where we have defined

$$G_n = -\frac{\tilde{E}_J}{2\hbar}\left[ Z_{p+n}^{\{N\}}(2i\vec{\Delta} \circ \vec{\alpha}) + Z_{p-n}^{\{N\}}(-2i\vec{\Delta} \circ \vec{\alpha}^*) \right], \tag{S8}$$

with $\vec{\alpha}$ taking the corresponding fixed point value. Note that $G_n$ does *not* appear in the fixed point expression, Eq. (S4), as the latter has a minus sign between the Z-functions.

The Jacobian is relatively sparse, with most elements zero in the symmetry preserving state: $G_n = 0$ except when $n = 0, \pm p, \pm 2p \ldots$. Where $G_n$ is nonzero, it can be expressed using Z-functions with dimension $M$ using the same mode reduction procedure as before.

### C. Block Diagonal Structure

The Jacobian matrix can be rearranged so that it takes a block diagonal form. The first stage involves separating out the terms that correspond to the resonant harmonics, formally we can write the Jacobian as $\overline{J} = \overline{J}_{\text{NH}} \oplus \overline{J}_{\text{H}}$ where $\overline{J}_{\text{H(NH)}}$ is a matrix of size $2M$ $(2(N-M))$ squared involving just the modes that are (not) resonant harmonics of $\omega_d = p\omega_1$; the direct sum ($\oplus$) notation builds the full Jacobian by placing the component block matrices corner to corner.

We can also carry out a simple reordering of the rows/columns to bring $\overline{J}_{\text{NH}}$ into a block diagonal form. These blocks are indexed by $k$ which labels the rotational symmetry subspace and runs from 1 to $(p-1)/2$ for $p$ odd and to $p/2$ for $p$ even. Block $k$ involves the modes $(k, p-k, p+k, 2p-k, 2p+k, 3p-k,...)$ terminating the sequence as appropriate given the number of modes $N$ [see Eqs. (7) and (8) in the main text], and takes the form

$$\overline{J}_k = \begin{pmatrix} \frac{\partial f_k}{\partial \alpha_k} & \frac{\partial f_k}{\partial \alpha^*_{p-k}} & \\ \frac{\partial f^*_{p-k}}{\partial \alpha_k} & \frac{\partial f^*_{p-k}}{\partial \alpha^*_{p-k}} & \\ & & \ddots \end{pmatrix} = \frac{\text{i}}{\hbar} \overline{\Delta}_k \begin{pmatrix} -G_0 & -G_{+p} & \\ +G_{-p} & +G_0 & \\ & & \ddots \end{pmatrix} \overline{\Delta}_k - (\gamma/2)\mathbb{I}. \tag{S9}$$

where $\overline{\Delta}_k$ is a diagonal matrix with elements $\Delta_k, \Delta_{p-k}, ....$ For odd $p$, the block of the Jacobian for modes which aren't resonant harmonics, $\overline{J}_{\text{NH}}$, consists of a direct sum of the $(p-1)/2$ different $k \neq 0$ blocks, $\overline{J}_k$, together with their complex conjugates, $\overline{J}_k^*$,

$$\overline{J}_{\text{NH}} = \bigoplus_{k=1}^{(p-1)/2} \left( \overline{J}_k \oplus \overline{J}_k^* \right), \tag{S10}$$

For even $p$, things are slightly more complicated as the block with $k = p/2$ is its own complex conjugate (it is similar to the band-edge mode in a crystal that lacks a counter propagating partner), and in this case

$$\overline{J}_{\text{NH}} = \left[ \bigoplus_{k=1}^{(p-1)/2} \left( \overline{J}_k \oplus \overline{J}_k^* \right) \right] \oplus \overline{J}_{p/2}. \tag{S11}$$

A continuous DTTSB transition occurs when an eigenvalue with a positive real part emerges in one of the $\overline{J}_k$ blocks within $\overline{J}_{\text{NH}}$. Once a subspace has developed an instability, the fixed point changes as amplitudes of the modes in that space become nonzero. These nonzero amplitudes can generate new couplings, breaking the strict division into symmetry sectors and affecting the amplitudes of modes which were originally in different blocks. This mechanism allows the instability to propagate from a single block to affect all of the modes in all of the blocks.

We searched for the critical $\lambda$ value at which an instability first occurs in $\overline{J}$ for a wide range of $N$ and $p$. The corresponding $k$ value of the unstable block is shown in Fig. S1. If the symmetry subspace of the instability is $k \neq 0$ then a continuous DTTSB transition can be identified. However, when the instability occurs in the $k = 0$ space the instability does not represent a DTTSB transition, but instead a bifurcation within the resonant harmonics. We did not identify the behavior of the resonant harmonics beyond the first bifurcation that they undergo and hence did not locate any continuous DTTSB transition in these cases.

### III. CRITICAL DRIVE STRENGTHS IN THE FEW-MODE REGIME: $N \leq p$

When $N \leq p$, the form of the Jacobian matrix, $\overline{J}$, is particularly simple with each $\overline{J}_k$ taking the form of a $2 \times 2$ matrix that allows analytic results to be derived. One finds that for $\Delta_n = \Delta_1/\sqrt{n}$ the first instability occurs in the $k = p - N$ subspace for $N < p$ or $k = 1$ when $N = p$, with a critical drive given by:

$$\lambda_c = \frac{\tilde{E}_J \Delta_1^2}{\hbar \gamma} = \begin{cases} \infty & 2N < p, \\ \sqrt{N(p-N)} & N < p \leq 2N, \\ \dfrac{\sqrt{p-1}}{2J_1'(2\Delta_p A_p)} & N = p. \end{cases} \tag{S12}$$

A continuous transition, occurring via a pitchfork bifurcation, is only possible in the presence of a quadratic term in the Hamiltonian[3,4]. This means that when the lowest order energy conserving processes produce 3 or more photons

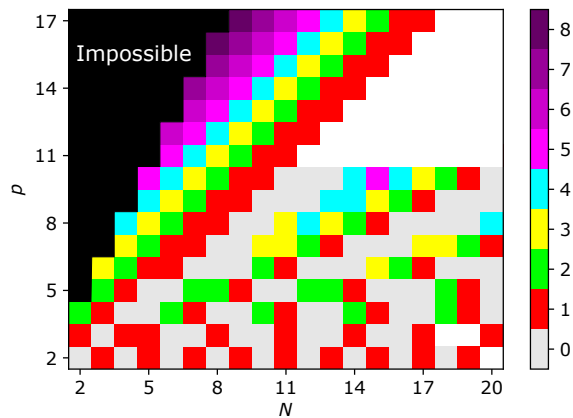

FIG. S1: Symmetry subspace, $k$, in which a continuous DTTSB transition emerges as a function of $N$ and $p$. Note that gray indicates no DTTSB transition found because an instability in the $k = 0$ space occurred first. In addition white indicates an $(N, p)$ combination that wasn't explored and black corresponds to cases where no transition is possible (see Sec. III).

only discontinuous transitions are possible. Hence, the first line of Eq. (S12): No continuous transition is possible if the drive frequency is more than twice that of the highest frequency mode.

For $N = p$ (a single resonant harmonic), we can derive a simpler expression for $\lambda_c$ assuming $p \gg 1$ and $\Delta_1 A_p/\sqrt{p} \ll 1$. Substituting Eq. (S12) into the fixed point expression[2] $\tilde{E}_J/(\hbar \gamma) = A_p^2/J_1(2\Delta_p A_p)$, provides an approximate equation for the amplitude of the resonant harmonic at the instability $\Delta_1 A_p \simeq 2$. Substituting this back into Eq. (S12) then leads to the approximate relation $\lambda_c \approx p\sqrt{p-1}/(p - 3/2)$. This approximation proves quite accurate for $p > 10$.

---



\end{document}